\documentclass[aps,twocolumn,prl,tightenlines,floatfix]{revtex4}
\usepackage[dvips]{graphicx}
\usepackage[english]{babel}
\usepackage{amsmath}
\usepackage{amssymb}
\usepackage{times}
\def\be{\begin{equation}}
\def\ee{\end{equation}}
\def\bea{\begin{eqnarray}}
\def\eea{\end{eqnarray}}

\newcommand{\ek}{\epsilon_{\mathbf{k}}}

\newcommand{\Ek}{{E^{}_{\mathbf{k}}}}

\newcommand{\mb}[1]{{\mathbf{#1}}}

\renewcommand{\text}[1]{{#1}}

\begin{document}

\title{Probing the Spectral Function Using Momentum Resolved Radio
  Frequency Spectroscopy in Trapped Fermi Gases}

\author{Qijin Chen, and K. Levin}

\affiliation{James Franck Institute and Department of Physics,
 University of Chicago, Chicago, Illinois 60637}

\date{\today}

\begin{abstract}
  We address recent momentum resolved radio frequency (RF) experiments
  on ultracold trapped Fermi gases of $^{40}$K. We show that momentum
  resolved RF probes provide measurements of the centrally important
  fermionic spectral function. They also serve to remove ambiguity
  plaguing the interpretation of momentum integrated RF experiments by
  establishing a clear dispersion signature of pairing.  We find that
  the temperature dependence of the spectral function is dramatic at
  unitarity, and, importantly, smooth from above to below $T_c$
  throughout BCS-BEC crossover.  This should be tested experimentally,
  given widespread predictions of first order behavior.
\end{abstract}

\maketitle

Ultracold Fermi gases undergo BCS to Bose-Einstein condensation (BEC)
crossover via a Feshbach resonance which is tuned with variable magnetic
field. Study of BCS-BEC crossover has been argued \cite{LeggettNature}
to be important for understanding high temperature superconductivity.
These ultracold Fermi gases and their extensions to optical lattices are
viewed as quantum simulators of some of the most important problems in
condensed matter physics. However, due to the extreme small size of the
Fermi gas cloud and the ultra low temperatures, the available
experimental probes are limited by comparison with their condensed
matter counterparts. Of particular importance are measurements of the
fermionic spectral functions, $A(\mathbf{k},\omega)$ for which angle
resolved photoemission spectroscopy (ARPES) has been so powerful in
electronic systems.  It is clear that counterpart experimental probes of
$A(\mathbf{k},\omega)$, in the ultracold gases are crucial to progress.

Recent experiments on $^{40}$K from the JILA group \cite{Jin6} have now
demonstrated that it is possible to measure these spectral functions
using momentum resolved RF pairing gap spectroscopy over a range of
magnetic fields throughout the BCS-BEC crossover.  There is a
substantial advantage of using $^{40}$K over the more widely studied
$^6$Li since, for the usual Feshbach resonance around 202 G, there are
no nearby competing resonances to introduce complications from final
state interactions.  In this Letter we address the theoretical basis
behind these measurements and demonstrate that this momentum resolved RF
technique has, in principle, the same capabilities as that of ARPES.  We
then show that this technique can be used to shed light on the recent
controversy \cite{Ketterlepairsize} about the origin of double peak
structure seen in previous \cite{Grimm4} momentum-integrated RF spectra,
as to whether it has to do with pairing or bound state effects.  Both
our calculations and the measurements in Ref.~\cite{Jin6} of the
momentum resolved RF spectral weight support the interpretation that a
generic double peak structure in $^{40}$K is associated with paired
atoms near the trap center and unpaired atoms near the trap edge.

Importantly, we find that, near unitarity, the spectral function has a
significant temperature variation. As in ARPES measurements of the
temperature dependent spectral function these experiments can provide
evidence for the presence of a normal state gap or pseudogap above $T_c$
at unitarity.  Indeed, a central focus of the cuprate literature has
been to address the evolution of angle resolved photoemission
experiments from above to below $T_c$.  This focus provides the
motivation for the present work, which addresses the counterpart to
these experiments in the cold Fermi gases. Our work is based on a
formalism that has been successfully applied to to ARPES measurements
\cite{ChenArcs} in the high temperature superconductors.  The
corresponding physical picture of the pseudogap which opens at a
temperature $T^* > T_c$ is that it arises from a stronger-than-BCS
attraction leading to preformed, normal state pairs.  The famous BCS
paper highlights an additional challenge which must be met by theory:
namely, explaining a smooth or \textit{second order phase transition} at
$T_c$.  In the presence of BCS-BEC crossover we find that the transition
should also be second order. This is at odds \cite{Strinati8,Zwerger}
with widespread predictions of first order behavior in the literature.

In addressing RF spectroscopy we can ignore final state effects
\cite{ourpaper} for the particular resonance under consideration. In the
recent JILA RF experiments \cite{Jin6} the momentum of state 3 atoms is
measured using time-of-flight imaging, in conjunction with 3D
distribution reconstruction.  Since the 3D gas in a single trap is
isotropic, detailed angular information is irrelevant.  In a homogeneous
system, the contributions to the RF current from state 2 to state 3 are
of two types corresponding to transitions associated with the breaking
of 1-2 pairs in the condensate or with the promotion of already existing
thermal fermionic excitations to level 3.
In RF transitions, a dominantly large fraction ($\Omega_L$) of the
photon energy $h\nu$ is used to excite the atoms in hyperfine level 2 to
another internal state of the atom, corresponding to level 3, where
$\Omega_L$ is the photon energy needed when the atoms in level 2 are free.
By contrast, in the ARPES case, there is an "emission" process in that
the photon energy is converted into the electron kinetic energy so that
the electrons are knocked out of the crystal.

\begin{figure}
  \includegraphics[width=3.0in,clip] {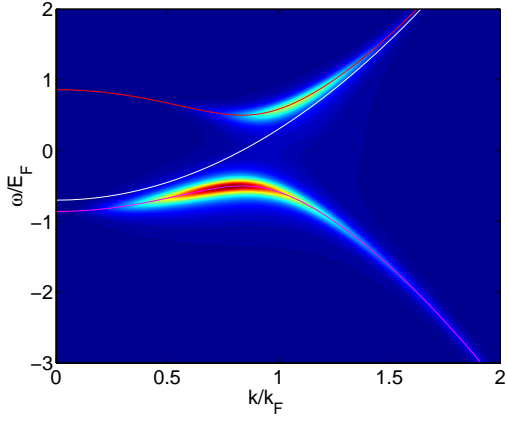}
  \caption{Contour plot of the occupied spectral intensity
    $A(k,\omega)f(\omega) k^2/2\pi^2$ at unitarity in the homogeneous
    case at $T/Tc \approx 1.9$.  The three lines in the contour plots
    correspond to the two quasiparticle dispersions $\pm \Ek$ and the
    free atom dispersion, respectively, where we take $\mu = 0.7E_F$,
    $\Delta = 0.5E_F$, $T=0.4T_F$, and $\gamma = 0.1E_F$. 
}
\label{fig:1}
\end{figure}

In the many-body theoretical derivation, it is convenient to choose the
Fermi level of state 2 as zero energy, so that we define
$\Omega_L^\prime$ as the overall energy level splitting between the
chemical potentials for atoms in level 3 and level 2. When pairing is
present, the Green's function $G(K)$ for this state is gapped, and the
Green's function $G_3^{(0)} (K)$ for state 3 is free. We then have for
the momentum-resolved RF current
\bea I(\mathbf{k}, \delta\nu)
  &=&  \frac{|T_k|^2}{(2\pi)^2} \int d\omega' A(\mathbf{k}, \omega'-\Omega)
  A_3(\mathbf{k}, \omega') \nonumber\\
 &\times&  [f(\omega)-f(\omega'+\Omega_L^\prime)]\nonumber\\ 
  &=&\left. \frac{|T_k|^2}{2\pi} A(\mathbf{k}, \omega)
  f(\omega)\right|_{\omega=\ek -\delta\nu} 
\label{eq:RFcurrent}
 \eea
 where $K\equiv (\mathbf{k}, i\omega_n)$ is the four-momentum,
 and $T_{\mathbf{k}}$ is the transition matrix
 element, which we will take to be a constant and then set to unity.
 This expression has been previously derived in Ref.~\cite{heyan},
 except there was also a momentum integral.  We have also assumed that
 state 3 is essentially empty [$f(\omega'+\Omega_L^\prime) = 0$], and
 has a free dispersion $A_3(\mathbf{k},\omega) = 2\pi
 \delta(\omega-\ek+\mu_3-\mu)$.  Note here $\delta\nu = \Omega
 -\mu+\mu_3 = h\nu -\Omega_L$ is the RF detuning and $\omega=\ek
 -\delta\nu$ is the energy of state 2 measured with respect to the Fermi
 level, where $\ek = k^2/2m -\mu$, $\mu$ and $\mu_3$ are the chemical
 potentials of atoms in state 2 and 3, respectively, and we take $\hbar
 =1$. The last line of Eq.~(\ref{eq:RFcurrent}) is formally identical to
 the expression for ARPES signal intensity. In this way, it should be
 clear that momentum-resolved RF is equivalent to ARPES. Both measure
 the fermionic spectral function.

 Beginning with Refs.~\onlinecite{Janko} and \onlinecite{Maly1}, there
 have been studies of the spectral function associated with BCS-BEC
 crossover which show that, beyond the BCS regime, there is a spectral
 gap or pseudogap.  Related work has also been performed by other groups
 \cite{Strinaticuprates}.  Detailed numerical calculations \cite{Maly1}
 of the related normal state self energy (called $\Sigma_{pg}$) have led
 to a rather simple BCS-like form, which has been shown
 phenomenologically \cite{Normanarcs} to be quite successful for
 describing the normal phase of the cuprates.  Following our earlier
 body of work \cite{Reviews,heyan,ChenArcs} we presume that the self
 energy for a short-range interaction consists of a term arising from
 the condensed, $\Sigma_{sc}$, as well as noncondensed pairs
 ($\Sigma_{pg}$). We write $\Sigma({\bf k}, \omega) = \Sigma_{pg}({\bf
   k}, \omega ) + \Sigma_{sc}({\bf k}, \omega )$, so that

\begin{eqnarray}
  \Sigma_{pg}(\mb{k},\omega) &=& 
  \frac{\Delta_{pg}^2}{\omega+\ek+i\gamma} -i\Sigma_0 \\
\label{SigmaPG_Model_Eq}
\Sigma_{sc}(\mb{k},\omega) &=&
\frac{\Delta_{sc}^2}{\omega +\ek} \:.
\label{SigmaSC}
\end{eqnarray}
Here $\gamma \ne 0$ reflects a phenomenological broadening term,
associated with a finite life time for incoherent pairs, and $\Sigma_0$
represents an ``incoherent'' background contribution, including that from
the particle-hole channel.
$\Sigma_{sc}$ is associated with long-lived, condensed Cooper pairs and
so it is of the same form as $\Sigma_{pg}$ but without the broadening.
With this self energy in the Green's functions, the resulting spectral
function, $A(\mb{k},\omega)=-2\,\mbox{Im}\, G(\mb{k},\omega+i0)$ can be
readily determined.  There is, moreover, a well defined procedure for
computing the two contributions (from the condensed and noncondensed
pairs) to the excitation gap which we do not repeat here.  This
formalism has been applied in the cold Fermi gas experiments
\cite{JS2,Torma2,heyan,heyan2} as well as in the cuprate literature
\cite{Chen4,ChenArcs}.

In Fig. 1, we present an intensity map as a function of the single
particle energy and the wave vector $k$ for the states which contribute
to the RF current in a homogeneous system.  The yellow, red regions of
the figure indicates those pairs of points $(E,k)$ where the greatest
number of contributing atoms lie.  The temperature here is chosen to be
relatively high, around $1.9 T_c$ in order to have appreciable
contributions to the RF current from pre-existing thermal fermionic
excitations.  The intensity map indicates upward and downward dispersing
contributions.  These correspond, to a good approximation, to the two RF
transitions to state 3 from state 2 with dispersions ($\Ek + \mu$) and
($-\Ek + \mu$) relative to the bottom of the band. Here $\Ek = \sqrt{
  \ek ^2 + \Delta ^2}$ is the quasiparticle excitation energy, and
$\Delta^2 = \Delta_{sc}^2 + \Delta_{pg}^2$ defines the total gap
$\Delta$, with $\Delta_{sc}=0$ above $T_c$.  The width of the peaks in
this contour plot comes exclusively from the incoherent terms $\gamma$
and $\Sigma_0$.  One can see that the bulk of the current even at this
high temperature is associated with the pair states which are broken in
the process of the RF excitation.  This figure describes in a conceptual
way, how this intensity map can be used to compare with a broadened
BCS-like form for the spectral function, which fits very well the two
branches shown in the figure corresponding to upward and downward
dispersing curves.  From the intensity map, one can determine the
spectral gap by fitting the spectral function or the dispersion to a
broadened BCS-like form, as has been done in Ref.~\cite{Jin6}.

\begin{figure*}
\centerline{
\includegraphics[height=1.7in,clip]{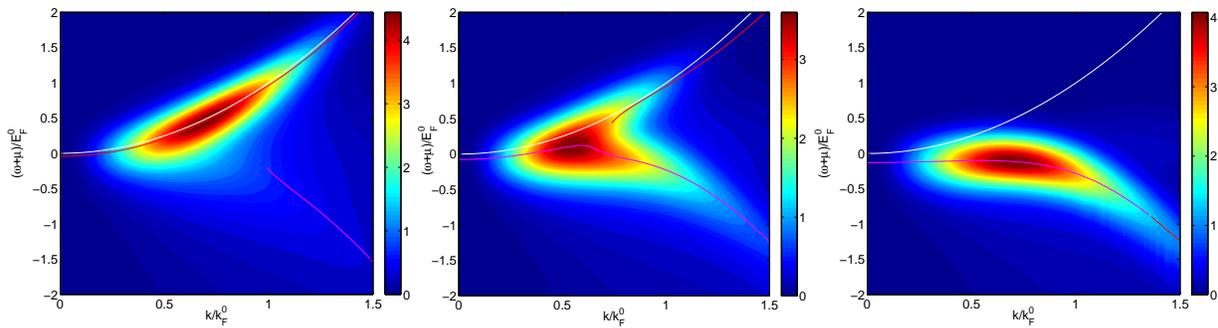}
}
\caption{Contour plots of the occupied spectral intensity
  $A(k,\omega)f(\omega) k^2/2\pi^2N$ at unitarity for (a) $T/T_c =
  1.35$, (b) 0.8, and (c) 0.1, from left to right. Here $\Sigma_0 = 0.25
  E_F^0$ and $\gamma = 0.25(T/T_c) E_F^0$ at the trap center, and both
  scale with local density as $r$ varies, $T_c \approx 0.27T_F^0$. The
  results were convoluted with a Gaussian broadening curve of width
  $\sigma = 0.22E_F^0$, whose value is taken from Ref.~\cite{Jin6}. As
  $T$ decreases, the spectral weight shifts from a free atom peak
  (upward dispersing curve) to a paired atom peak (downward dispersing
  curve).  Here the lines indicate the loci of the peaks in the EDCs.}
\label{fig:2}
\end{figure*}

We stress that the above calculations are for the homogeneous case and
it is important to extend them to include the effects of a trap, where
the density and gap reach maximum at the trap center and vanish at the
trap edge.  This is done within a local density approximation
(LDA). Once the trap is included the simple analogy between the
electronic ARPES experiments and momentum resolved photoemission
spectroscopy is invalidated. While $\delta\nu$ is constant across the
sample, the single particle excitation energy $\omega =
k^2/2m-\mu(r)-\delta\nu$ will necessarily bear the inhomogeneity effect
of the spatial dependence of $\mu(r)$; after time of flight imaging, one
cannot distinguish atoms from different radii $r$. It is thus more
convenient to measure the single particle energy from the band bottom,
$\omega+\mu(r) = k^2/2m-\delta\nu$, which is constant across the trap
for given $k$. Nevertheless, many of the central features survive.
While in a homogeneous system there will be two branches as described
above, in a trap, there is a third branch which appears as well. This
corresponds to essentially free atoms at the trap edge which will
contribute significantly to the RF current. This contribution dominates
that of the thermally excited quasiparticles at both low and high
$T$. It is this branch which is also upward dispersing and makes it
rather difficult to see the contributions from pre-existing thermally
broken pairs. With improved experimental resolution, this momentum
resolved RF probe can in principle distinguish these two contributions
by their dispersion.

The results of the LDA-based calculations for the intensity map plots
are shown in Fig.~2 for the unitary gas in a trap, for different
temperatures, calculated with $\Sigma_0 = \gamma = 0.25E_F^0$, and
convoluted with an experimental Gaussian line of width $\sigma = 0.22
T_F^0$. 
The vertical axis represents the trap averaged single
particle energy, $(\omega+\mu)/E_F^0$, where $E_F^0 = (k_F^0)^2/2m $ is
the global Fermi energy in the non-interacting limit. The temperatures
are $T/T_c = 1.35$ (above $T_c$), 0.8 (slightly below $T_c$), and 0.1
(well below $T_c$), from left to right.  At the highest $T$ the central
notable feature is a single upward dispersing curve which fits the free
particle dispersion; the spectral weight is dominated by contributions
from atoms at the trap edge. This dispersion can be readily
differentiated from that associated with pre-existing thermally broken
pairs which varies as $\Ek + \mu$ and of course, depends on the
distribution of energy gaps $\Delta(r)$.

As the temperature decreases a second (downward dispersing) branch
becomes evident. The spectral weight shifts to the lower branch
gradually. The intensity map first bifurcates and eventually becomes
dominated by the lower branch at very low $T$, where essentially all
atoms are paired. Here the lower branch is associated with the
breaking of a condensate pair. It, moreover, contains trap averaging 
effects.
In the absence of final state effects,
this can be used to determine the pairing gap.
%
%
The separation of the two peaks can be difficult to discern until
sufficiently high values of $k$. 


That there are no abrupt changes at the superfluid transition
distinguishes this theoretical scheme from others (e.g.,
Refs.~\cite{Strinati8,Zwerger}) in the literature.  Just as for
the high temperature superconductors, there are no first order
transitions as the system evolves from above to below $T_c$.  
The behavior at unitarity is in some sense smoother
than in the strict BCS theory, since we find that a pairing gap is
present above $T_c$, which shows up as a bifurcation in the intensity
plots.  There are indications \cite{Jin6} from the experimental momentum
resolved RF plots that the behavior evolves smoothly from the normal to
the superfluid phase, but detailed quantitative analyses to rule in or
out possible first order transitions are lacking.
In Fig. 3A we plot the energy distribution curves (EDC) for a series of
momenta $k$ in a unitary trapped Fermi gas at $T/T_c = 1.1$ as a
function of single particle energy.  Here we have taken a somewhat
larger intrinsic and instrumental broadening than in Fig.~1. These
parameters are seen to optimize semi-quantitative agreement with the
data. Thus the EDC in Fig.~3A can be compared with their counterparts
plotted in Fig.~4 in Ref.~\cite{Jin6}. At small $k$ there is a single
peak which separates into two at higher $k$ values.  The data are not
sufficiently smooth to be assured that two peaks are visible at larger
$k$, but they are consistent with this picture.  As before, we attribute
the separation at higher $k$ to the fact that the upward and downward
dispersing curves become most well separated in this regime.  Better
agreement is expected with improved signal-to-noise ratio in the
data. Moreover, including detailed incoherent background contributions
in the fermionic self-energy may also improve the agreement with
experiment.

\begin{figure}
\centerline{\includegraphics[height=1.5in,clip]{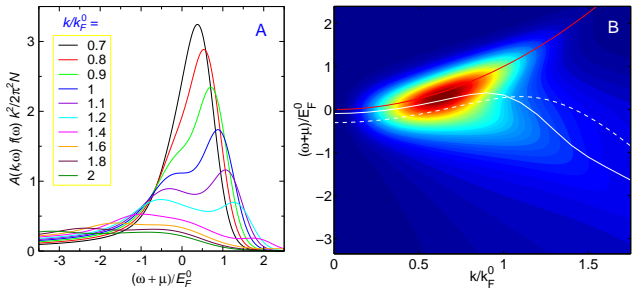}}
\caption{(A) Energy distribution curves for a series of momenta $k$ as
a function of single particle energy, and (B) corresponding occupied
spectral intensity map, in a unitary trapped Fermi gas at $T/T_c =
1.1$. Here $\Sigma_0 = 0.35 E_F^0$ and
$\gamma = 0.38 E_F^0$ at the trap center, and $\sigma = 0.3
E_F^0$. In (B), the red curves represent the free atom dispersion,
while the white solid and dashed curves are the quasiparticle
dispersion obtained theoretically and experimentally \cite{Jin6},
respectively, via fitting the EDCs with a single Gaussian
distribution.}
\label{fig:3}
\end{figure}

Figure 3B presents the corresponding spectral intensity map.  The white
dashed curve represents a fit of the experimental peak dispersion
\cite{Jin6} while the solid white curve is the theoretical counterpart.
Here, as in experiment we have fit the EDC to a \textit{single} Gaussian
peak. The comparison between the two white curves shows
semi-quantitative consistency.  Moreover, both the solid and dashed
white curves can be well fit to the BCS dispersion involving $\Ek$, as
was originally proposed in Ref.~\cite{Janko}.  Indeed Fig.~3B seems to
capture the essential results shown in Fig.~3a in Ref.~\cite{Jin6}.
With higher experimental resolution it should be possible to obtain more
direct information about the mean gap size. In these experiments on
$^{40}$K, final state effects should be of less importance in
complicating the interpretation. The fact that the experiments were done
near $T_c$ suggests that there is a sizable pseudogap at and above $T_c$
at unitarity.

We turn finally to the BEC side of resonance in Fig.~4.  Here the
intensity map is presented for $1/k_F^0 a = 1.1$ at a temperature $T/T_c
= 1.3$, where $T_c \approx 0.3 T_F^0$. Even at this relatively high
temperature there is virtually no weight in the free atom peak,
associated with atoms at the trap edge.  All the atoms are bound into
pairs and there is only one downward dispersing peak in the plot.  This
behavior can be contrasted with that shown in Fig.~3c of
Ref.~\cite{Jin6} where there is a free atom contribution, which has been
attributed to unpaired atoms out of chemical equilibrium with the pairs.
In addition the downward dispersing peak in the left panel is rather
sharp. In contrast, the large peak width in Ref.~\cite{Jin6} seems to
suggest incoherent contributions to the fermionic self energy, which may
have to do with chemical non-equilibrium. The underlying theory here
presumes that there are still well defined quasi-particles on the BEC
side of resonance.  We present in Fig.~3b the case in which we have
considered a substantially larger intrinsic broadening $\gamma \approx
\Delta$ to represent some of the qualitative
features of the experimental data.

\begin{figure}
\centerline{
\includegraphics[height=1.7in,clip]{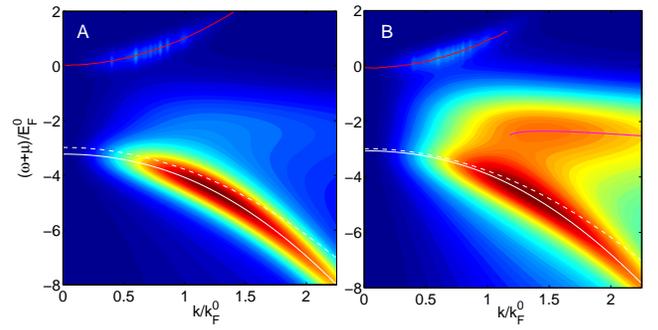}
}
\caption{Occupied spectral intensity maps at $1/k_F^0a = 1.1$ in the BEC
case and $T/T_c = 1.3$. The broadening parameters at the trap center
are $\Sigma_0 = 0.35 E_F^0$ and $\gamma = 0.45 E_F^0$ for the left
panel and $\Sigma_0 = 1.0 E_F^0$ and $\gamma = 1.3 E_F^0$ for the
right panel. For both, $\sigma = 0.25 E_F^0$. Coherent sharp
quasiparticle peaks can be seen in the left panel. Larger broadening in the
right panel makes the coherent peak less pronounced, mimicking
possible effects of chemical non-equilibrium. The red curve represents
free atom dispersion, and the solid and dashed white curves represent
the theoretical and experimental quasiparticle dispersion,
respectively.}
\label{fig:4}
\end{figure}

In conclusion, we have shown that momentum resolved RF spectroscopy can
be used to measure the centrally important spectral function of
ultracold Fermi gases.  Other spectroscopic experimental tools have been
proposed in the literature \cite{Georges}, although currently there are
no counterpart experimental studies.  We show that our calculated
spectral intensity maps are in semi-quantitative agreement with
experiments on $^{40}$K.  Because they establish the actual dispersion
associated with the RF current contributions, these experiments remove
ambiguity associated with whether pairing contributions might be
confused with final bound state effects \cite{Ketterlepairsize}.  Both
our theory and the experiments in Ref.~\cite{Jin6} reveal that a double
peak structure reminiscent of that seen in previous RF spectra
\cite{Grimm4} is indeed associated with paired atoms at the trap center
and unpaired atoms at the trap edge. At unitarity, we show that there is
substantial temperature variation in the spectral weight distribution.
Theoretically, the spectral intensity maps and quasiparticle dispersion
evolve smoothly with temperature across $T_c$. Whether first order
transitions exist, as some have predicted, will require further
experiments.

This work is supported by NSF PHY-0555325 and NSF-MRSEC Grant 
No.~DMR-0213745. We thank Debbie Jin, Jayson Stewart and John Gaebler for 
many useful communications.


\end{document}